\documentclass[twocolumn,a4paper,amsmath,amssymb,showpacs,prb,superscriptaddress]{revtex4-1}
\usepackage{amsmath}
\usepackage{bm}
\usepackage{graphicx}
\usepackage{epstopdf}
\usepackage[normalem]{ulem}
\usepackage{color}

\begin{document}

\title{First-principles study of magnetic properties in Fe-ladder
compound BaFe$_2$S$_3$}

\author{Michi-To Suzuki}
\affiliation{RIKEN, Center for Emergent Matter Science, 2-12-1 Hirosawa, Wako, Saitama 351-0198, Japan}
\author{Ryotaro Arita}
\affiliation{RIKEN, Center for Emergent Matter Science, 2-12-1 Hirosawa, Wako, Saitama 351-0198, Japan}
\author{Hiroaki Ikeda}
\affiliation{Department of Physical Sciences, Ritsumeikan University, 1-1-1 Nojihigasi, Kusatsu, Shiga 525-8577, Japan}

\date{\today}

\begin{abstract}
  We study the magnetic, structural, and electronic properties of the recently discovered
 iron-based superconductor BaFe$_2$S$_3$ based on density
 functional theory with the generalized gradient approximation.
 The calculations show that the magnetic alignment in which the spins are
 coupled ferromagnetically along the rung and antiferromagnetically along 
 the leg is the most stable in the possible
 magnetic structure within an Fe-ladder and is
 further stabilized with the periodicity characterized by the wave
 vector {\bf Q}=($\pi$,$\pi$,0),
 leading to the experimentally observed magnetic ground state. 
 The magnetic exchange interaction between the Fe-ladders creates a tiny
 energy gap, whose size is in excellent agreement with the experiments.
 Applied pressure suppresses the energy gap and leads to an
 insulator-metal transition.
 Finally, we also discuss what type of orbitals can play crucial
 roles on the magnetic and insulator-metal transition.
\end{abstract}

\pacs{}

\maketitle

\section{introduction}
The discovery of superconductivity in fluorine doped LaFeAsO has
stimulated intensive studies of iron-based
superconductors.~\cite{Kamihara2008}
In addition to the high superconducting transition temperature and the strong
flexibility for the constituent atoms, similarities of magnetic and
structural properties with the high-$T_c$ cuprate superconductors are
suggestive to understand the pairing mechanism.~\cite{Scalapino2012}
For instance, the parent material undergoes antiferromagnetic transition
under ambient pressure, and the superconducting transition appears in
the vicinity of the antiferromagnetic phase by doping carriers or by
applying pressure.

\begin{figure}[b]
\begin{center}
\includegraphics[width=1.0\linewidth]{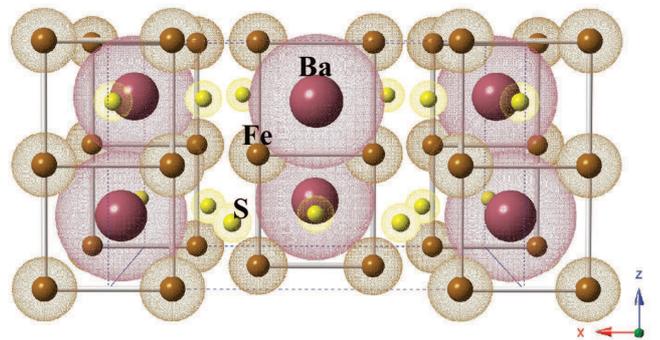}
\caption{(Color online) Crystal structure of BaFe$_2$S$_3$. }
\label{Fig:CrystalStruct}
 \end{center}
\end{figure}

Recently, superconductivity in BaFe$_2$S$_3$ has been discovered under
pressure with the maximum $T_c$ $\sim$ 14K.~\cite{Takahashi2015}
 Investigation of the magnetic and electronic properties in relation
 to the structural property in this ladder system is expected to provide
 a new path to understand the high $T_c$ superconductivity in the
 iron-based superconductors.
  The quasi one-dimensional Fe-ladder formation of BaFe$_2$S$_3$, as
  shown in Fig.\ \ref{Fig:CrystalStruct}, exhibits a striking difference from the
 conventional iron-based superconductors, which have two-dimensional Fe
 networks, but rather
reminds us of the superconducting two-leg ladder cuprate Sr$_{14}$Cu$_{24}$O$_{41}$.~\cite{Uehara1996}
 While the magnetic structure of BaFe$_2$S$_3$ has a marked
 similarity to conventional iron-based superconductors as will be
 discussed below,
 BaFe$_2$S$_3$ has a semiconducting ground state as contrasted to
 the metallic ground states of other iron-based superconductors. 

 A number of structurally related compounds of BaFe$_2$S$_3$, i.e. {\it
 A}Fe$_2${\it X}$_3$ ({\it A}=K, Rb, Cs, or Ba and X = S, Se, Te), have
 been discovered.~\cite{Hong1972, Klepp1996, Reissner2006}
 In particular, the block magnetism of BaFe$_2$Se$_3$ has attracted
 increasing attention as the novel magnetic structure among iron-based materials
 and has been intensively studied both theoretically~\cite{Li2014, Luo2014, Dagotto2013, Luo2013, Lv2013, Medvedev2012} and
 experimentally.~\cite{Komedera2015,Mourigal2014,Lovesey2014, Nambu2012,
 Caron2012, Krzton2011, Caron2011, Saparov2011, Lei2011, Hong1972}
On the other hand, the physical properties of BaFe$_2$S$_3$ are reported only
 recently in association with the discovery of the high-$T_c$ superconducting transition.~\cite{Takahashi2015} 
 BaFe$_2$S$_3$ crystallizes in an orthorhombic crystal structure
 whose symmetry belongs to space group $Cmcm$ (No. 63), which is
 the same as those of CsFe$_2$Se$_3$ and KFe$_2$Se$_3$~\cite{Caron2012, Du2012} but is
 different from $Pnma$ (No. 62) of BaFe$_2$Se$_3$~\cite{Hong1972}.
 While the block magnetism in BaFe$_2$Se$_3$ is characterized by {\bf
 Q}=($\pi$,$\pi$,$\pi$), the periodicity of the stripe magnetism in BaFe$_2$S$_3$ is
 characterized by {\bf Q}=($\pi$,$\pi$,0). The same stripe-type
 magnetism with BaFe$_2$S$_3$ is also observed in CsFe$_2$Se$_3$ and  KFe$_2$Se$_3$.

 The local magnetic moment of iron is reported as 1.2$\mu_B$ in
 BaFe$_2$S$_3$,~\cite{Takahashi2015} which is quite smaller than 1.8$\mu_B$ of
 CsFe$_2$Se$_3$~\cite{Du2012} and 2.8$\mu_B$ of
 BaFe$_2$Se$_3$~\cite{Nambu2012}, and  the N{\' e}el temperatures of 120K,
 175K, and 255K for  BaFe$_2$S$_3$, CsFe$_2$Se$_3$, and BaFe$_2$Se$_3$,
 respectively, are corresponding to the size of the magnetic moments.
 The activation energy of BaFe$_2$S$_3$ is reported to be less than 0.1
 eV,~\cite{Goenen2000, Reiff1975} which is again smaller than
 0.13 eV and 0.178 eV reported for
 BaFe$_2$Se$_3$~\cite{Nambu2012, Lei2011} and 0.34 eV for CsFe$_2$Se$_3$.~\cite{Du2012}

\begin{figure}[tb]
\begin{center}
\includegraphics[width=0.7\linewidth]{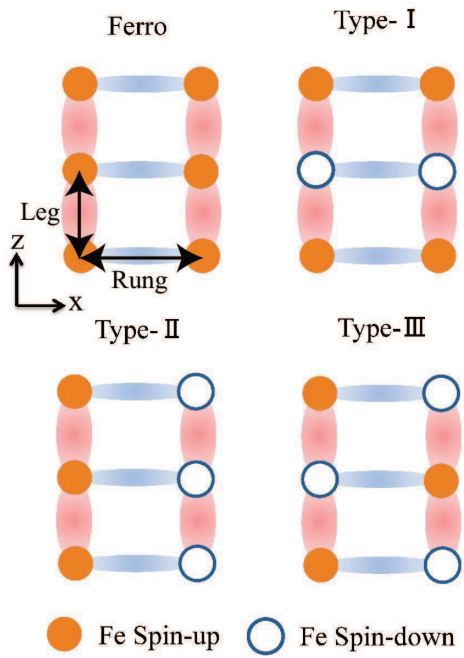}
\caption{(Color online) Possible inequivalent magnetic configurations within the Fe-ladder.}
\label{Fig:SpinConfig}
 \end{center}
\end{figure}

 The first-principles study for iron-based superconductors have provided
 fruitful information concerning the structural, magnetic, and electronic properties.~\cite{Ishibashi2008,
 Ishibashi2008a, Mazin2008a, Mazin2008, Yin2008, Yildirim2008, Yildirim2009}
 For BaFe$_2$S$_3$, we recently derived a low energy effective Hamiltonian of BaFe$_2$S$_3$ based on
 a first-principles approach.~\cite{Arita2014}
 In this paper, we investigate the magnetic and electronic properties in
 relation to the structural property under pressure by using the first-principles
 calculations based on the generalized gradient approximation (GGA).
 We show that the stripe spin-ladder magnetic order, which is
 refered to type-I magnetic configuration
 in Fig.\ \ref{Fig:SpinConfig} with the ordering wave vector {\bf Q}=($\pi$,$\pi$,0), is the most stable
 magnetic structure over the investigated range of pressure.
 We also discuss how the electronic structure changes under pressure and
 undergoes a metal-insulator transition, which is necessary for the
 superconducting transition.

\begin{figure}[tb]
\begin{center}
\includegraphics[width=1.0\linewidth]{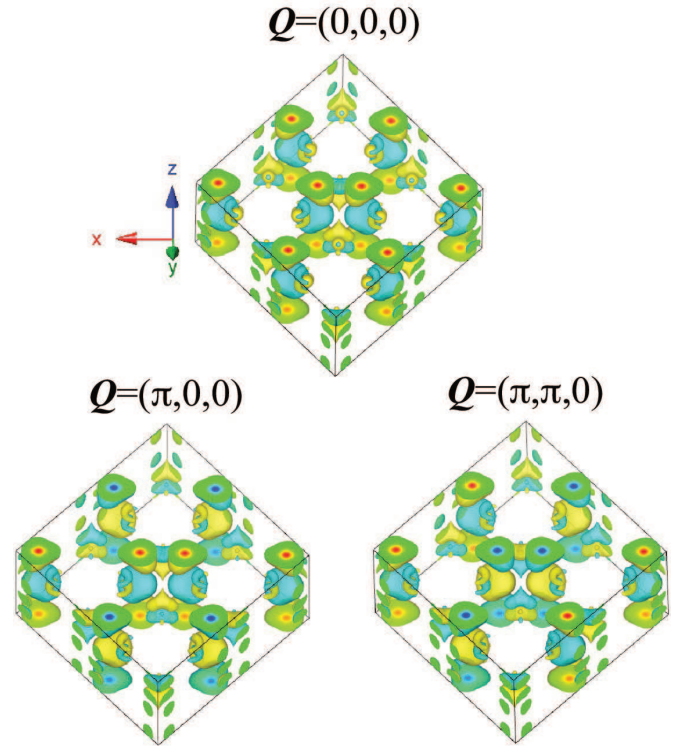}
\caption{(Color online) Computed spin distribution of the type-I magnetic order characterized by ${\bf
 Q}$=(0,0,0), ($\pi$,0,0), ($\pi$,$\pi$,0). The yellow (blue) surface is
 an isosurface of positive (negative) spin density.}
\label{Fig:SpinDistribution}
 \end{center}
\end{figure}
 
\section{method}
The first-principles calculations are implemented in the VASP program
code with projector augmented-wave (PAW) method using the
exchange-correlation functional proposed by Perdew, Burke, and Ernzerhof.~\cite{Perdew1996}
 We systematically explored all the possible spin configurations within
 the Fe-ladder, as shown in Fig.\ \ref{Fig:SpinConfig},  for the
 characteristic wave vectors {\bf Q}=(0,0,0), ($\pi$,0,0),
 ($\pi$,$\pi$,0), setting the 2$\times$2$\times$1 super cell,
 corresponding to the magnetic unit cell for the periodicity characterized by {\bf Q}=($\pi$,$\pi$,0).
 The $k$-point mesh is set as the (2,2,2) for the self-consistent
 calculations and (8,8,8) for plotting the density of states (DOS) for
 the calculations of the super cell.
 The calculations were performed with the experimental lattice constants
 $a$=8.78 \AA, $b$=11.23 \AA, $c$=5.29 \AA\ for ambient pressure, and the
 pressure effect are considered as change in the lattice constants
 observed experimentally.~\cite{Hirata2015} 
 The atomic relaxation is performed for the sulfur atoms within the $x$-$y$
 plane, starting from the initial atomic configuration as the
 experimental values i.e. Ba(4c) (0.0,0.686,0.25), Fe(8e)
(0.154,0.0,0.0), S(4c) (0.0, 0.116, 0.25), and S(8g) (0.208,
0.378, 0.25), using the experimental lattice constants.~\cite{Hirata2015}
 The energy cutoff for the plane waves is set as 500 eV through the calculations.
 The convergence conditions for the total energies are set as
 10$^{-5}$ eV for electronic self-consistent loop and 10$^{-2}$ eV for the ionic relaxation.
\begin{widetext}
\begin{center}
\begin{table*}[h]
 \begin{tabular}{ccccccc} \hline
                         &   \multicolumn{6}{c}{Ferro} \\
                      & \multicolumn{2}{c}{Experimental structure
      ($P$=0)}&\multicolumn{2}{c}{Optimized structure ($P$=0)} &
  \multicolumn{2}{c}{Optimized structure ($P$=16.27 GPa)} \\
                         & Total energy (eV)  & Fe-moment ($\mu_B$) & Total
  Energy (eV)  & Fe-moment ($\mu_B$)& Total energy (eV)  & Fe-moment
  ($\mu_B$) \\
    {\bf Q}=(0,0,0)      & -291.006           & 1.28   & -293.901 & 0.00 & -293.296 & 0.00 \\
 {\bf Q}=($\pi$,0,0)     & -290.786           & 1.08   & -293.901 & 0.00 & -293.295 & 0.00 \\
 {\bf Q}=($\pi$,$\pi$,0) & -291.050           & 1.03   & -293.901 & 0.00 & -293.296 & 0.00 \\
  \hline
                         &   \multicolumn{6}{c}{Type-I} \\
                      & \multicolumn{2}{c}{Experimental structure
      ($P$=0)}&\multicolumn{2}{c}{Optimized structure ($P$=0)} &
  \multicolumn{2}{c}{Optimized structure ($P$=16.27 GPa)}\\
                         & Total energy (eV)  & Fe-moment ($\mu_B$) & Total
  Energy (eV)  & Fe-moment ($\mu_B$) & Total energy (eV)  & Fe-moment
  ($\mu_B$) \\
     {\bf Q}=(0,0,0)     & -296.009           & 2.26   & -296.778  & 2.08 & -293.558 & 1.42 \\
 {\bf Q}=($\pi$,0,0)     & -295.967           & 2.27   & -296.742  & 2.08 & -293.632 & 1.10 \\
 {\bf Q}=($\pi$,$\pi$,0) & -296.045           & 2.27   & -296.809  & 2.07 & -293.672 & 1.40 \\
  \hline
                         & \multicolumn{6}{c}{Type-II} \\
                      & \multicolumn{2}{c}{Experimental structure
      ($P$=0)}&\multicolumn{2}{c}{Optimized structure  ($P$=0)} &
  \multicolumn{2}{c}{Optimized structure ($P$=16.27 GPa)} \\
                         & Total energy (eV)  & Fe-moment ($\mu_B$)  &
  Total energy (eV)  & Fe-moment ($\mu_B$) & Total energy (eV)  & Fe-moment
  ($\mu_B$) \\
  {\bf Q}=(0,0,0)        & -294.964           & 2.55              & -295.245 & 2.42 & -293.296 & 0.00 \\
 {\bf Q}=($\pi$,0,0)     & -294.953           & 2.49              & -295.310 & 2.31 & -293.296 & 0.00 \\ 
 {\bf Q}=($\pi$,$\pi$,0) & -294.653           & 2.49              & -295.043 & 2.30 & -293.296 & 0.00 \\
  \hline
                         & \multicolumn{6}{c}{Type-III} \\
                      & \multicolumn{2}{c}{Experimental structure
      ($P$=0)}&\multicolumn{2}{c}{Optimized structure  ($P$=0)} &
  \multicolumn{2}{c}{Optimized structure ($P$=16.27 GPa)} \\
                         & Total energy (eV)  & Fe-moment ($\mu_B$) & Total
  Energy (eV)  & Fe-moment ($\mu_B$) & Total energy (eV)  & Fe-moment
  ($\mu_B$) \\
  {\bf Q}=(0,0,0)        & -294.402           & 2.15              & -295.314 & 1.87 & -293.295 & 0.15 \\
 {\bf Q}=($\pi$,0,0)     & -294.259           & 2.14              & -295.182 & 1.89 & -293.296 & 0.00 \\
 {\bf Q}=($\pi$,$\pi$,0) & -294.281           & 2.15              & -295.195 & 1.86 & -293.295 & 0.14 \\
\hline
 \end{tabular}
    \caption{Total energies and spin moments calculated for the
 different spin configurations (see text).}
\label{Tab:TEnergy}
\end{table*}
 \end{center}
\end{widetext}

\section{results}
 Firstly, we discuss the stable magnetic structure of BaFe$_2$S$_3$.
  The magnetic distributions of type-I magnetism with
 different ${\bf Q}$ are shown in Fig.\ \ref{Fig:SpinDistribution}.
 Calculated total energies and Fe magnetic moments are shown in table\ \ref{Tab:TEnergy}.
  As shown in the table, the calculations predict the most stable
 magnetic structure as the type-I magnetic alignment, in which the
  spins are coupled ferromagnetically along the rung and
  antiferromagnetically along the leg,
 with the periodicity characterized by the ${\bf Q}$=($\pi$,$\pi$,0),
 which corresponds to the experimentally observed magnetic alignment.
  The energy difference is mainly dominated by the spin configuration
  within the Fe-ladders, changing the total energy with the order of 1 eV, 
  and the different ${\bf Q}$ structure change it with the order of 0.1 eV. 

 Relaxation of sulfur atoms somewhat reduces the magnetic moments from
 those for the experimental atomic position for the type-I, II, and III
magnetic states and completely suppresses it for FM spin alignment.
 Remarkably, the type-I magnetic configuration is the most stable magnetic
 structure within an Fe-ladder in
 spite of the smaller magnetic moments compared to those of type-II.
 This result indicates that the atomic configuration strongly affects to
 stabilize the type-I magnetism through coupling with the magnetic
 moments.
 The predicted Fe's magnetic moments 2.07$\mu_B$ of the type-I magnetic order is
 considerably larger than the experimental value 1.3$\mu_B$, and this
 kind of overestimation for Fe's moments was also reported for LaFeAsO.
 ~\cite{Yin2008,Ishibashi2008,Ishibashi2008a,Mazin2008,Mazin2008a,Hansmann2010,Toschi2012}

\begin{figure}[tb]
\begin{center}
\includegraphics[width=0.8\linewidth]{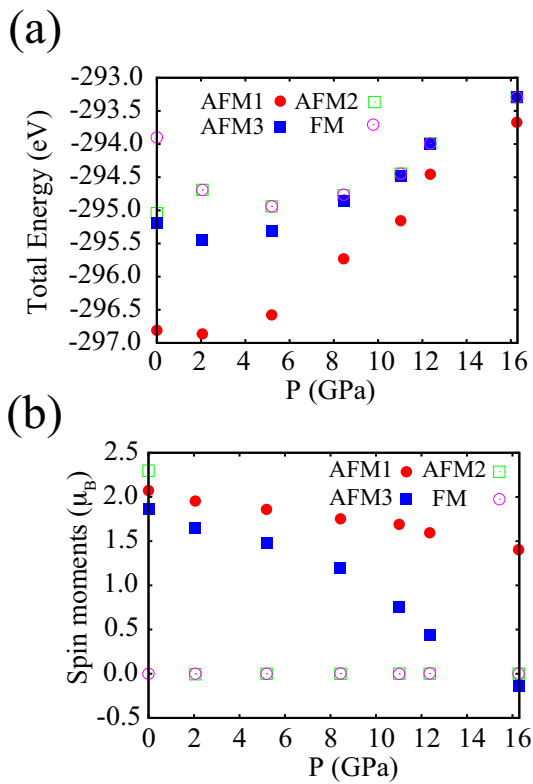}
\caption{(Color online) Pressure dependence of (a) total energy and (b) magnetic
 moments for all possible magnetic structures characterized by ${\bf Q}$
=($\pi$,$\pi$,0).}
\label{Fig:TEneMagMom}
 \end{center}
\end{figure}

 We also investigated the magnetic states under pressure by using the
 experimental lattice constants with performing relaxation of sulfur atoms. 
 As shown in table\  \ref{Tab:TEnergy}, whereas the magnetic moments
 of the type-I magnetic states are not much reduced at $P$=16.27 GPa, those
 of the other magnetic states are strongly suppressed under the high pressure. To examine 
 the pressure dependence of the magnetic states in further detail, we show in
 Fig.\ \ref{Fig:TEneMagMom} the detailed pressure
 dependence of the total energies and Fe-spin moments for the magnetic
 states with the periodicity {\bf Q}=($\pi$,$\pi$,0). From the figure we
 can find the clear difference in the pressure dependence for all of the magnetic
 structures, though the type-I, II, and III magnetism exhibit the same order of
 the magnetic moments at ambient pressure.
 The pressure effect strongly suppresses the magnetic moments of the type-II magnetism,
 in which the magnetic moment vanishes already at $P$=2 GPa and
 gradually suppress the magnetic moments of the type-III magnetism, leading to
 the nonmagnetic state around the 16 GPa.
 On the other hand, the pressure dependence of the magnetic moment for the
 type-I magnetism is rather weak and leave the magnetic order stable
 even under high pressure.

\begin{figure}[tb]
\begin{center}
\includegraphics[width=0.8\linewidth]{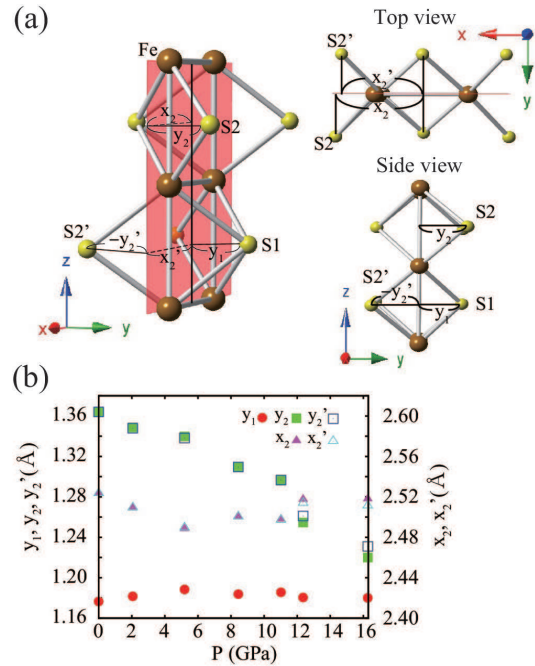}
\caption{(Color online) (a)Definitions of sulfur's relative positions
 and (b) the calculated values for the type-I magnetic states. The red
 surface sheet in figure (a) indicates a surface on which the Fe-ladder lies.}
\label{Fig:SulfurPosi}
 \end{center}
\end{figure}

 The optimized sulfur positions, whose coordinates
 are defined in Fig.\ \ref{Fig:SulfurPosi}, are $y_{\rm S1}$=1.18 \AA, 
 $x_{\rm S2}$=$x_{\rm S2'}$=2.52 \AA, and $y_{\rm S2}$= $y_{\rm
 S2'}$=1.36 \AA\ while the experimental positions are $y_{\rm S1}$=1.30 \AA, 
 $x_{\rm S2}$=$x_{\rm S2'}$=2.57 \AA, and $y_{\rm S2}$=$y_{\rm
 S2'}$=1.37 \AA. The deviation between the calculation and
 the experiment for the S1 position is somewhat larger than that for the
 S2 atom, yet the
 deviations are about 1\% at most for the lattice constants.
 We also show the pressure dependence of the sulfur's atomic positions in
 Fig.\ \ref{Fig:SulfurPosi}, in which the positions of sulfurs are measured from
 the Fe-ladder as defined in Fig.\ \ref{Fig:SulfurPosi} (a). 
 Interestingly, from Fig.\ \ref{Fig:SulfurPosi} (b), significant shifts
 in the sulfur positions are only observed for the sulfurs out of the
 Fe-ladders (S2 and S2'), and the
 positions of sulfurs located between the ladder's legs (S1) does not strongly
 depend on the pressure.
 Note that S2 and S2' in Fig.\ \ref{Fig:SulfurPosi} are equivalent in the
 non-magnetic crystal structure but are inequivalent in the
 magnetic state. The calculations indeed show small deviation of the
 two sulfur positions above $P\sim$12 GPa as shown in Fig.\ \ref{Fig:SulfurPosi} (b).

\begin{figure}[h]
\begin{center}
\includegraphics[width=0.7\linewidth]{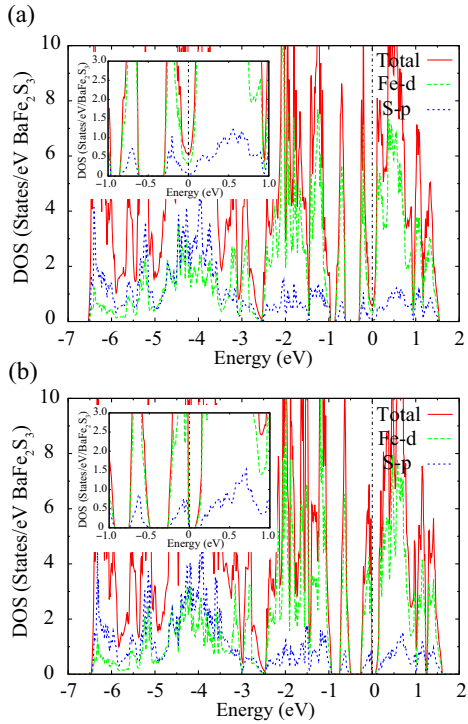}
\caption{(Color online) Density of states of Type-I AFM magnetic ordered states with
 the periodicity characterized by {\bf Q}=(0,0,0) and by {\bf
 Q}=($\pi$,$\pi$,0). Insets are the enlarged views around the
 Fermi levels.}
\label{Fig:dos_Qdiff}
 \end{center}
\end{figure}
\begin{figure}[h]
\begin{center}
\includegraphics[width=0.8\linewidth]{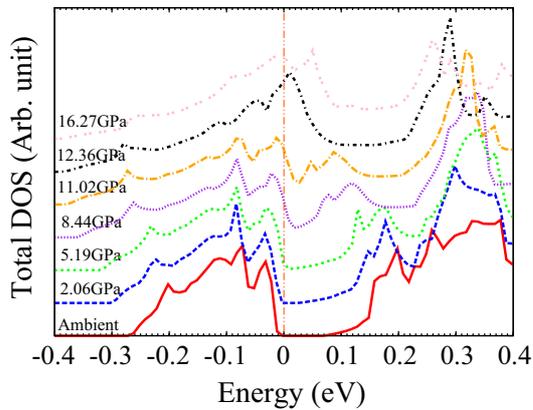}
\caption{(Color online) Pressure dependence of density of states for
 type-I AFM magnetic ordered states with the periodicity of {\bf Q=($\pi$,$\pi$,0)}.}
\label{Fig:dos_pressure}
 \end{center}
\end{figure}
\begin{figure}[b]
\begin{center}
\includegraphics[width=1.0\linewidth]{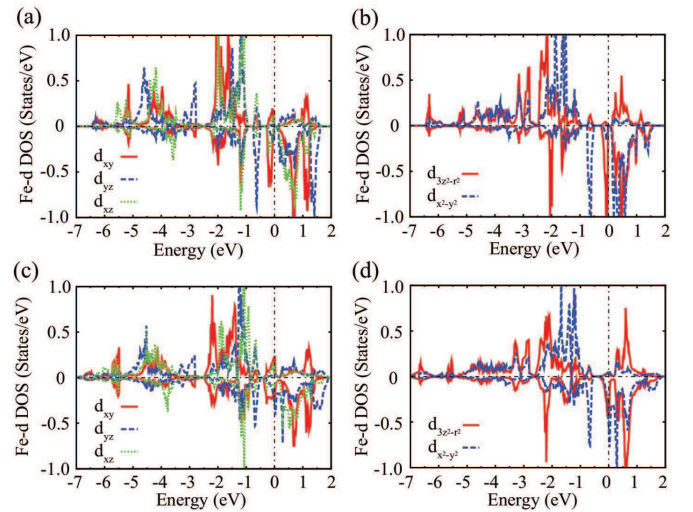}
\caption{(Color online) Density of states of Fe-$d$ orbitals
 for Type-I ({\bf Q}=($\pi$,$\pi$,0)) AFM states for
 $P$=0 and 12.36 GPa.}
\label{Fig:dos_orbital}
 \end{center}
\end{figure}

 Finally, we discuss the electronic structure of BaFe$_2$S$_3$.
 Experimentally, BaFe$_2$S$_3$ is semiconductor with the small energy
 gap of 0.06-0.07 eV.~\cite{Goenen2000, Reiff1975}
 As shown in Fig.\ \ref{Fig:dos_Qdiff}, the energy gap of the
 BaFe$_2$S$_3$ is not produced with the {\bf
 Q}=(0,0,0) periodicity but produced with the periodicity of {\bf Q}=($\pi$,$\pi$,0)
 with the size 0.07 eV for the type-I magnetism, which is in excellent
 agreement with the experimental evaluations.
 This result indicates that the magnetic exchange coupling between the
 Fe ladders is crucial to create the tiny energy gap.
 Figure \ref{Fig:dos_pressure} shows pressure dependence of total DOS for the
 {\bf Q}=($\pi$,$\pi$,0) type-I antiferromagnetic states around the Fermi
 level.  
 The pressure strongly suppresses the energy gap and leads to a metallic
 state around $P\sim$5 GPa.
 The peak structure of DOS located below the Fermi level comes close to
 the Fermi level as the pressure increases and passes over the Fermi level
 at $P\sim$12 GPa. As a result, pressure strongly enhances the DOS at the
 Fermi level which takes its maximum at $P\sim$12 GPa.
 The orbital resolved DOS of Fe-$d$ orbitals for the pressure
 $P$=0 and 12.36 GPa is shown in Fig.\ \ref{Fig:dos_orbital}.
  The large difference in the structure of the DOS for the majority and
 minority spin indicates that all of the Fe-$d$ orbitals contribute to the
 magnetic order. More precisely, contribution for
 the spin-moment of each $d$ orbital calculated from the spin-DOS is 0.37 (0.28) for $d_{xy}$, 0.39
 (0.30) for $d_{yz}$, 0.34 (0.27) for
 $d_{3z^2-r^2}$, 0.47 (0.36) for $d_{xz}$, and 0.49 (0.37) for $d_{x^2-y^2}$,
 leading to the total Fe-$d$ moment 2.06 (1.58), at $P$=0 (12.36 GPa).
 The larger magnetic moments of the $d_{xz}$ and $d_{x^2-y^2}$ orbitals compared
 to other $d$-orbitals show the primary contribution of these two orbitals for
 the magnetism, which is consistent to our recent studies showing that low-energy
 effective Hamiltonians can be constructed from these two orbitals.~\cite{Arita2014}
  The $d_{3z^2-r^2}$ orbital mainly contributes to the DOS peak below the
 Fermi level at ambient pressure but applied pressure suppresses it as
 in Fig.\ \ref{Fig:dos_orbital} (b) and (d). 
 Alternatively, the $d_{yz}$ and the $d_{x^2-y^2}$ orbitals develop with
 applied pressure and contribute to the maximum DOS peak at Fermi level
 at $P\sim$12 GPa. Although further investigation is necessary to reveal
 the orbital contribution to the superconducting transition, the interplay
 of these Fe-$d$ orbitals should be crucial for the high-$T_c$ transition temperature.

\section{Conclusion}
 We investigated the magnetic, structural, and electronic properties of
 BaFe$_2$S$_3$ based on the first-principles calculations. The GGA
 calculations predicted the stripe-type spin-ladder magnetism, whose magnetic
 periodicity is characterized by {\bf Q}=($\pi$,$\pi$,0), in the ground state.
 The calculations of the pressure dependence for the total energy and magnetic moments
 indicate that the most stable type-I magnetic structure is
 not sensitive against the applied pressure compared to other magnetic
 structure and leave the
 magnetic order stable even at high pressure as a result.
 The magnetic exchange interaction between the Fe ladders creates
 a tiny energy gap at ambient pressure and is therefore a crucial
 ingredient to reproduce the insulating magnetic ground state as
 observed experimentally. Applied pressure strongly
 suppresses the energy gap and leads to a metallic magnetic state,
 bringing the high DOS at the Fermi level due to the formation of the
 DOS peak structure around the Fermi level.
 The magnetic moment predicted at the ambient pressure is considerably larger
 than that observed experimentally. The discrepancy of the magnetic
 states may be improved with more advanced first-principles method
 considering the strong electron correlation in the iron-based compound,~\cite{Hansmann2010,Toschi2012}
 and more precise prediction of the magnetic states is our future issue.

\bibliographystyle{revtex}
\bibliography{Fe123Systems}

\end{document}